\documentclass[twocolumn,superscriptaddress,prl,longbibliography,floatfix]{revtex4-1}
\makeatletter
\newcommand*{\rom}[1]{\expandafter\@slowromancap\romannumeral #1@} 
\makeatother
\usepackage{gensymb} 
\usepackage{graphicx}
\usepackage{times}
\usepackage{dcolumn}
\usepackage{hyperref}
\usepackage{caption}
\usepackage{color}
\usepackage{float}
\usepackage{amsmath, amssymb}
\usepackage{courier}
\usepackage{ulem}

\newcommand\BNA[0]{BaNi$_2$As$_2$}

\hypersetup{colorlinks = true, linkcolor = blue, anchorcolor =blue, citecolor = blue, filecolor = blue, urlcolor = blue,
	pdfauthor=author}
\begingroup

\begin{document}


\title{Multi-condensate lengths with degenerate excitation gaps in BaNi$_2$As$_2$ revealed by muon spin relaxation study}
\author{Kaiwen Chen}
\affiliation{State Key Laboratory of Surface Physics, Department of Physics, Fudan University, Shanghai 200438, China}
\author{Zihao Zhu}
\affiliation{State Key Laboratory of Surface Physics, Department of Physics, Fudan University, Shanghai 200438, China}
\author{Yaofeng Xie}
\affiliation{Department of Physics and Astronomy, Rice University, Houston, Texas 77005, USA}
\author{Adrian D. Hillier}
\affiliation{ISIS Facility, STFC Rutherford Appleton Laboratory, Harwell Science and Innovation Campus, Didcot OX11 0QX, United Kingdom}
\author{James S. Lord}
\affiliation{ISIS Facility, STFC Rutherford Appleton Laboratory, Harwell Science and Innovation Campus, Didcot OX11 0QX, United Kingdom}
\author{Pengcheng Dai}
\affiliation{Department of Physics and Astronomy, Rice University, Houston, Texas 77005, USA}
\author{Lei Shu}
\email{leishu@fudan.edu.cn}
\affiliation{State Key Laboratory of Surface Physics, Department of Physics, Fudan University, Shanghai 200438, China}
\affiliation{Shanghai Research Center for Quantum Sciences, Shanghai 201315, People's Republic of China}

\begin{abstract}
The recently discovered (Ba,Sr)Ni$_2$As$_2$ family provides an ideal platform for investigating the interaction between electronic nematicity and superconductivity. Here we report the muon spin relaxation ($\mu$SR) measurements on \BNA. Transverse-field $\mu$SR experiments indicate that the temperature dependence of superfluid density is best fitted with a single-band $s$-wave model. On the other hand, the magnetic penetration depth $\lambda$ shows magnetic field dependence, which contradicts with the single-band fully-gapped scenario.  Zero-field $\mu$SR experiments indicate the absence of spontaneous magnetic field in the superconducting state, showing the preservation of time-reversal symmetry in the superconducting state. Our $\mu$SR experiments suggest that \BNA\ is a  fully-gapped multiband superconductor. The superconducting gap amplitudes of each band are nearly the same while different bands exhibit different coherence lengths. The present work helps to elucidate the controversial superconducting property of this parent compound, paving the way for further research on doping the system with Sr to enhance superconductivity.
\end{abstract}

\maketitle

\section{\boldmath{$\rm{\rom1}$}. INTRODUCTION} \label{sec:intro}
One of the core issues in quantum materials is how superconductivity interacts with the Fermi surface instabilities such as charge density wave (CDW) and nematic order (NO)~\cite{Fradkin2015,Keimer2015,Johnston2010}. Although the CDWs in cuprates have been discovered for nearly three decades~\cite{Nature1995}, their contribution to anomalous superconducting properties has not yet been thoroughly understood~\cite{RN2,Proust2019,Agterberg2020}. In the Fe-based superconductors, the nematic order causing $C_4$-rotational symmetry-breaking strongly intertwines with the spin density wave~\cite{Dai2015,Fernandes2014,Paglione2010}. By suppressing these orders with pressure or doping, the superconductivity appears and is maximumly enhanced near the nematic and/or magnetic quantum critical point (QCP)~\cite{Marel2003,Lederer2015}. However, the entanglement between plural electronic instabilities complicates our understanding of the relationship between enhanced superconducting pairing and such instabilities.

Enormous efforts have been made to elucidate the physics of Fe-based superconductors. In this process, many other new superconducting systems were discovered~\cite{Wang2012,Jiang2015,Gu2022}. Among them, the nickel-based \BNA, which shares the same high-temperature tetragonal structure with the intensively studied 122 iron arsenide superconductors~\cite{PfistererNagorsen+1980+703+704,Alireza2009,Mani2009,Gallais2008}, has attracted attention recently. Compared to its iron-analog, which does not superconduct without doping at ambient pressure, BaNi$_2$As$_2$ itself shows superconductivity below $T_{\rm{c}}$ $\sim$ 0.6~K~\cite{Ronning2008} and undergoes a structure transition into a triclinic phase near $T^*\sim$ 130 K~\cite{Sefat2009}. An increasing number of recent studies have revealed an exotic normal state, characterized by complex charge density instabilities~\cite{Lee2019,Lee2021,Guo2022} and dynamic nematic fluctuations~\cite{Yao2022}. Though early studies on electronic properties have led to conventional BCS superconductivity~\cite{Shein2009,Subedi2008}, substituting Ba with Sr can boost $T_{\rm{c}}$ from 0.6 K to 3.6 K with the suppression of structural transition to zero temperature, which indicates the possible nematic-enhanced superconductivity near the optimal doping~\cite{eckberg2020,Lederer2020}. Therefore, BaNi$_2$As$_2$ provides an ideal platform to study how nematic fluctuations itself can influence superconductivity as there is no magnetic order or spin fluctuations in the system. 

Up to now, most of the studies on \BNA~are focused on the driving force of structure transition or the origin of CDWs.
In fact, the nature of superconductivity of BaNi$_2$As$_2$ is still unclear. On the one hand, early $ab~initio$ band structure calculation~\cite{Subedi2008} and angle-resolved photoemission spectroscopy (ARPES) measurements~\cite{Zhou2008} have revealed the presence of multiple bands crossing the Fermi level. The S-shaped field dependence of thermal conductivity suggested the multiple nodeless superconducting gaps~\cite{Kurita2009}. On the other hand, both specific-heat and thermal conductivity studies exclude the appearance of a second gap~\cite{Kurita2009,Ronning2008}. For unconventional superconductors coexisting with several electronic instabilities, the multiband property is a crucial ingredient since it is closely related to the pairing mechanism~\cite{Kittaka2014}. Therefore, it is important to determine the superconducting property of BaNi$_2$As$_2$.

Muon-spin relaxation/rotation ($\mu$SR) is a powerful technique to study the gap symmetry of superconductors by providing direct information on magnetic penetration depth $\lambda$~\cite{Aidy2022,Sonier2004}. In this work, transverse-field (TF)-$\mu$SR experiments have been performed to study the temperature and magnetic field dependence of $\lambda$ of BaNi$_2$As$_2$. The temperature dependence of superfluid density $\rho_{\rm{s}}$ is well-fitted with a single-band $s$-wave function with zero temperature gap amplitude $\Delta_0=0.101(6)$ meV. On the other hand, the field dependence of muon spin relaxation rates is well-fitted with a two-band model. Furthermore, $\lambda$ showing a magnetic field dependence, contradicting with single-band fully-gapped scenario. Our TF-$\mu$SR results suggest that BaNi$_2$As$_2$ is a multiband fully-gapped superconductor with equal gap value but different coherence lengths in each band. The zero-field (ZF)-$\mu$SR experiments reveal that the time-reversal symmetry is preserved in the superconducting state of \BNA.
\section{\boldmath{$\rm{\rom2}$}. EXPERIMENTAL DETAILS} \label{sec:exp}

High-quality single crystals of \BNA\ were synthesized at the Rice University with the same procedure described in Ref~\cite{Ronning2008}. The single crystal X-ray diffraction (XRD) experiment suggested the high purity of our samples. 

The $\mu$SR experiments were carried out on the MuSR spectrometer at ISIS Neutron and Muon Facility, Rutherford Appleton Laboratory, Chilton, UK. Several pieces of single crystals are mounted on the high purity silver plate with diluted GE varnish. The experiments were carried out in temperatures ranging from 50 mK to 1.0 K for both ZF and TF $\mu$SR setup. TF-$\mu$SR experiments were performed under external magnetic fields from 5 mT to 40 mT. The sample was field-cooled down above $T_{\rm{c}}$ to base temperature in order to form an ideal vortex lattice. All the $\mu$SR spectra were collected upon warming and the data were analyzed with the \texttt{MUSRFIT} software package~\cite{MUSRFIT}.

\section{\boldmath{$\rm{\rom3}$}. RESULTS} \label{sec:res}

\subsection{\boldmath{A. } Zero-field muon spin relaxation} \label{sec: ZFuSR}

\begin{figure}[H]
	\begin{center}
		\includegraphics[width=0.45\textwidth]{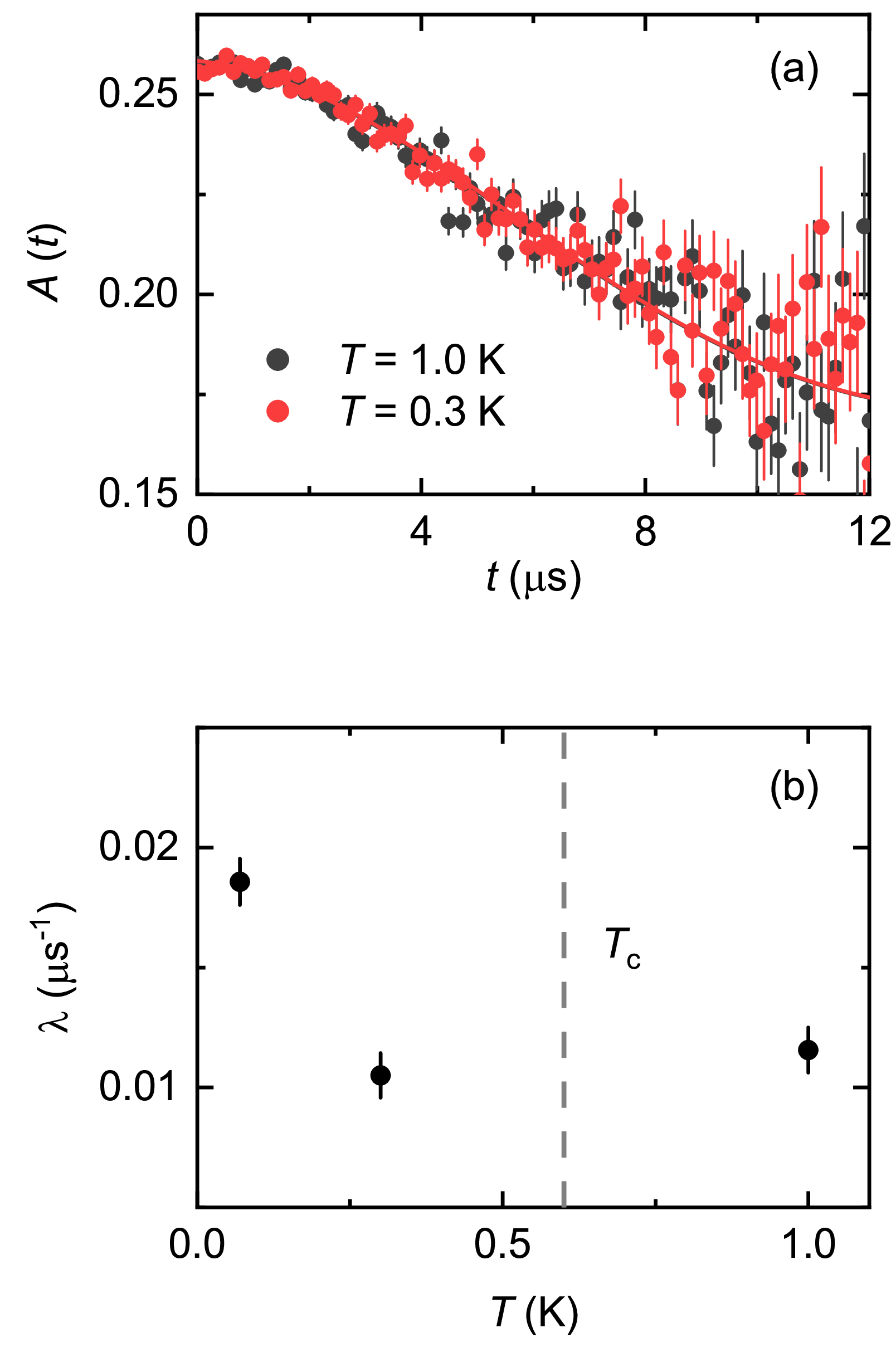}
		\caption{(a) ZF muon decay positron asymmetry $A(t)$, proportional to the muon spin polarization, measured in the normal state (black circles) and the superconducting state (red circles) respectively. The solid curves are the fits of data with Eq.~(\ref{eq:1}). (b) Temperature dependence of muon spin relaxation rate $\lambda$ down to 70 mK. The grey dashed line indicates the superconducting critical temperature $T_{\rm{c}}$.}
		\label{fig:Fig1}
	\end{center}
\end{figure}

The zero-field (ZF) $\mu$SR measurement was performed to figure out the magnetic ground state of BaNi$_2$As$_2$ and to examine whether the superconducting transition accompanied with an additional broken of time-reversal symmetry. Two representative ZF-$\mu$SR spectra measured above and below $T_{\rm{c}}$ are shown in Fig.~\ref{fig:Fig1}. No oscillations or loss of initial asymmetry were observed, suggesting the absence of both long-range and short-range magnetic order down to 70~mK. This is consistent with the previous neutron scattering study~\cite{Kothapalli2010}. ZF muon spin asymmetry spectra after subtracting the background signal $A_{\rm ZF}$ can be well fitted with a damped Gaussian Kubo-Toyabe (KT) function~\cite{Yaouanc2011MuonSR}:
\begin{eqnarray}
	\label{eq:1}
	A_{\rm {ZF}}(t)=A_0(T)G_{\rm {KT}}(\sigma_{\rm{ZF}},t)e^{-\lambda t}.
\end{eqnarray}
The KT term describes a Gaussian distribution of randomly oriented static (or quasi-static) local fields at muon sites with distribution width $\delta B_{\rm G}= \sigma_{\rm ZF}/\gamma_{\mu}$, where $\gamma_{\mu}= 2\pi \times 135.53$~MHz/T is the muon gyromagnetic ratio~\cite{Hayano1979}:
\begin{eqnarray}
	\label{eq:2}
	G_{\rm{KT}}(\sigma_{\rm{ZF}},t)=\frac{1}{3}+\frac{2}{3}(1-\sigma_{\rm{ZF}}^2t^2)\cdot\exp(-\frac{1}{2}\sigma_{\rm{ZF}}^2t^2),
\end{eqnarray}
The exponential term represents additional electronic depolarization rate~\cite{Yaouanc2011MuonSR}. $\sigma_{\rm{ZF}}$ is found to show little temperature dependence and is fixed to the average value of $\sigma_{\rm{ZF}}=0.122\ \rm{\mu s^{-1}}$ during the fitting. The temperature dependence of $\lambda$ is displayed in Fig.~\ref{fig:Fig1}(b). Though $\lambda$ grows at the lowest temperature, whose origin is not clear based on the present data, it does not show obvious change crossing $T_{\rm{c}}$. The current ZF-$\mu$SR result suggests there is no spontaneous time-reversal symmetry breaking exists with superconducting transition.

\subsection{\boldmath{B.} Transverse-field muon spin relaxation} \label{sec:TFuSR}
To elucidate the pairing symmetry and the band character of superconductivity in BaNi$_2$As$_2$, systematic temperature-dependent $\mu$SR experiments were performed under several transverse magnetic fields from 5 mT to 40 mT. The lower critical field $H_{c1}$ of \BNA\ is estimated with~\cite{Brandt2003}: 
\begin{eqnarray}
	\label{eq:3}
	H_{c1}=H_{c2}\frac{\ln{\kappa}+0.497}{2\kappa^{2}}.
\end{eqnarray}
Considering $\kappa\approx11$ estimated from normal state thermal conductivity~\cite{Kurita2009}, Eq.~(\ref{eq:3}) gives $H_{c1}\approx$ 1.3 mT. Although this is only an estimation, the applied magnetic field of 5 mT is large enough to form the uniform vortex lattice in \BNA.

\begin{figure}[H]
	\begin{center}
		\includegraphics[width=0.45\textwidth]{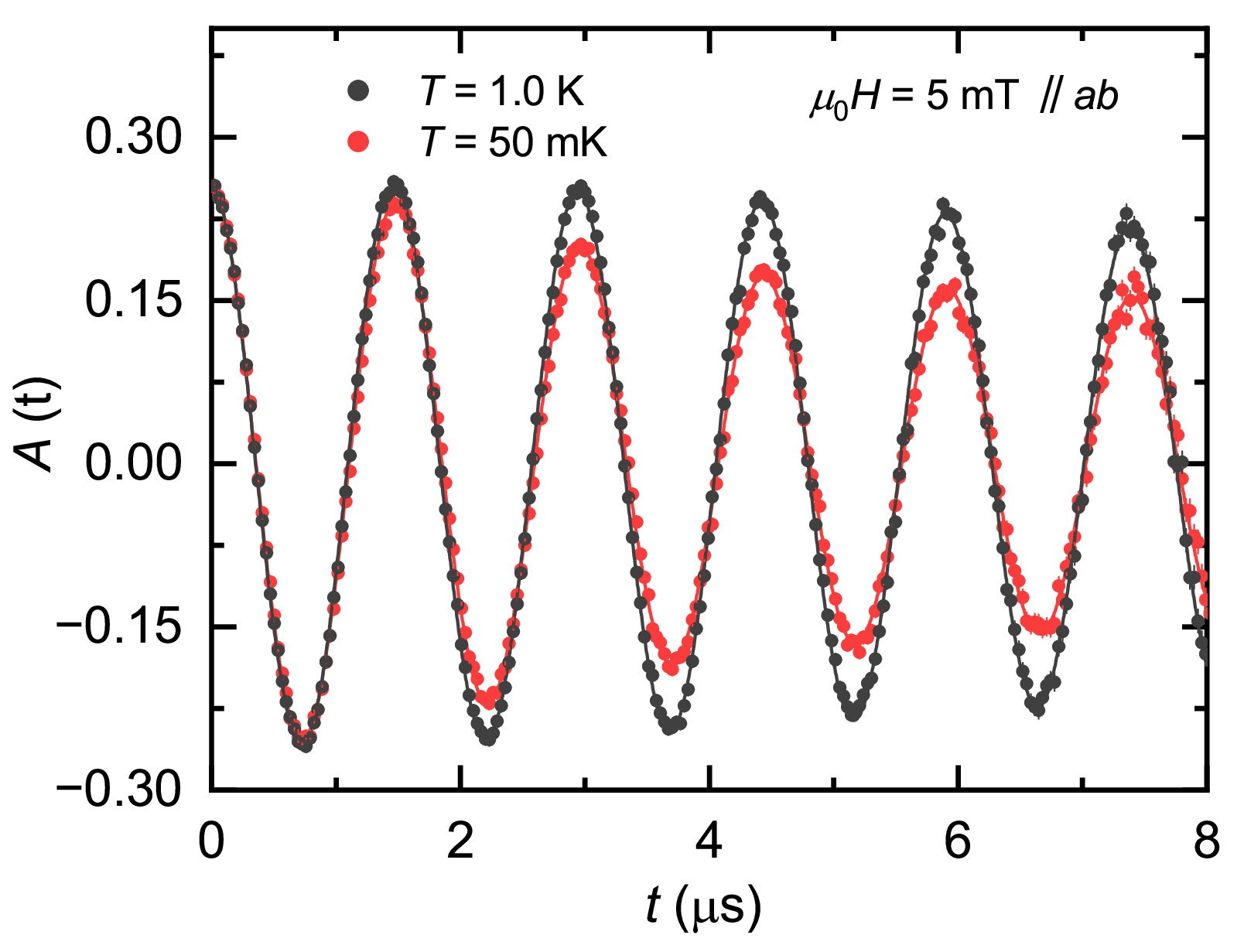}
		\caption{TF-$\mu$SR asymmetry spectra measured at $\mu_0H=5~{\rm mT}$. The black and red circles represent the asymmetry spectra in the normal state and in the superconducting state respectively.  The solid curves are the fits of data with Eq.~(\ref{eq:4}). }
		\label{fig:Fig2}
	\end{center}
\end{figure}

Two representative TF-$\mu$SR spectra above and below $T_{\rm{c}}$ are shown in Fig.~\ref{fig:Fig2}. The damping of asymmetry is enhanced after entering the superconducting state, indicating the bulk superconductivity of our sample. The superconducting volume fraction can be estimated from the difference of long-time spectra~\cite{Combley1985}. However, the relaxation of asymmetry spectra is relatively slow in \BNA, and its asymmetry spectra failed to fully relax even when the data points started to diverge (after 8 $\mu$s, not shown). Nevertheless, based on the asymmetry spectra at 8 $\mu$s, the lower limit of superconducting volume fraction can be estimated to be about 75$\%$. 

The TF-$\mu$SR data is analyzed with the function:
\begin{multline}
	\label{eq:4}
	A(t)/A_{0}=[(1-f)\exp(-\sigma^2t^2/2)\cos(\gamma_{\mu}\mu_0 H_{\rm{int}}+\phi)\\+f\exp(-\sigma_{\rm{bg}}^2 t^2)\cos(\gamma_{\mu}\mu_0 H_{\rm{bg}}+\phi)],
\end{multline}
where $A_0$ stands for the initial asymmetry; $f$ represents the proportion of the background signal and $(1-f)$ represents the proportion of sample signal; $\sigma$ is the Gaussian relaxation rate, which is related to the Gaussian field distribution in the sample; $\mu_0H_{\rm{int}}$ is the average field sensed by muons stopping in the sample; $\sigma_{\rm{bg}}$ is the relaxation rate of muon stopping in the silver, which is temperature independent and is fixed during the fitting~\cite{Bueno2011,Zhang2018}; $\mu_0H_{\rm{int}}$ is the field sensed by muon stopping in the background; and $\phi$ is the shared initial phase.  

\begin{figure}[H]
	\begin{center}
		\includegraphics[width=0.45\textwidth]{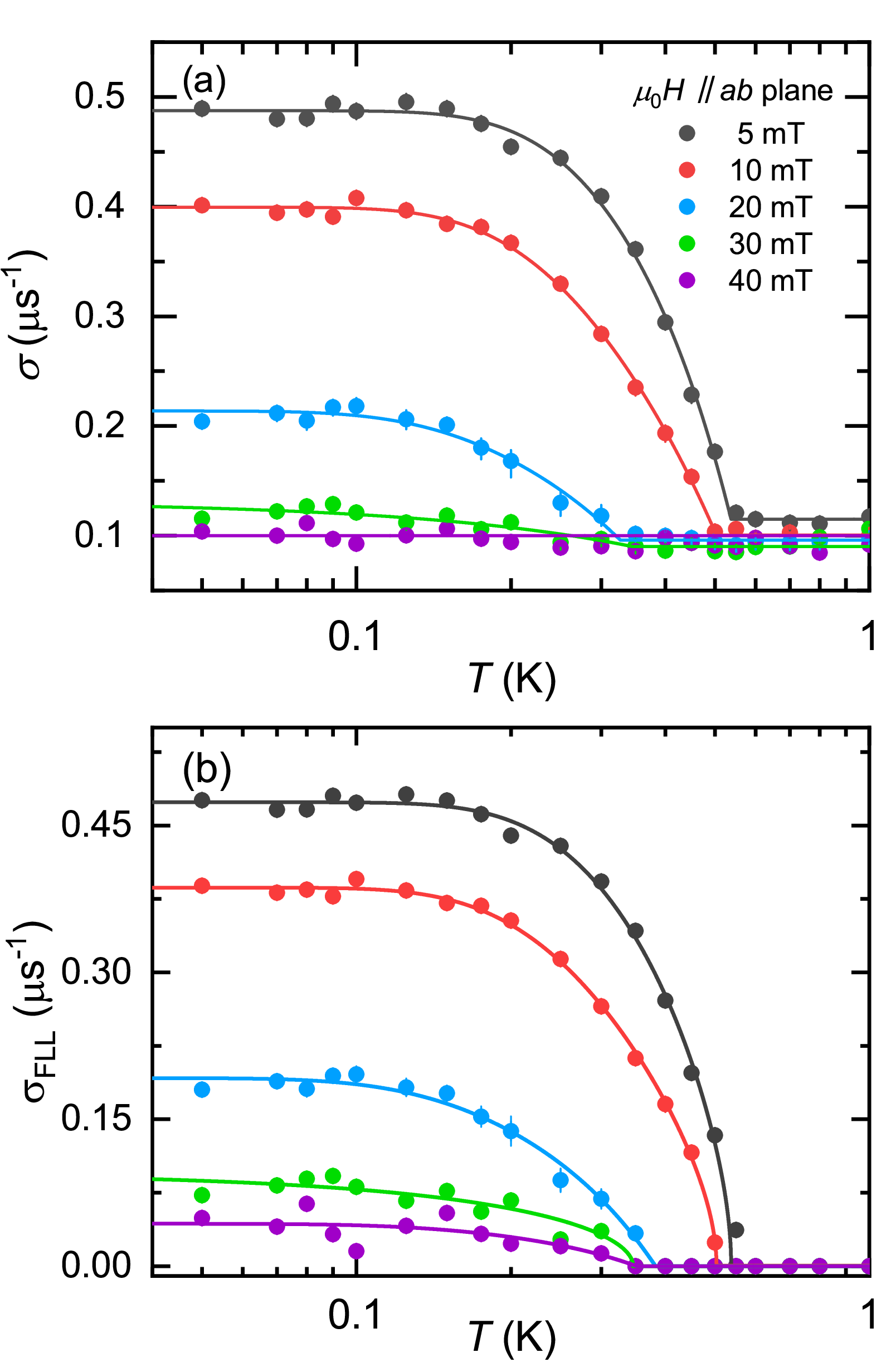}
		\caption{Temperature dependence of muon spin relaxation rates (a) $\sigma$ and (b) $\sigma_{\rm{FLL}}$ under different external fields respectively. The solid lines are guides for eyes.}
		
		\label{fig:Fig3}
	\end{center}
\end{figure}

The extracted relaxation rates $\sigma$ with Eq.~(\ref{eq:4}) are displayed in Fig.~\ref{fig:Fig3}(a). For $T > T_{\rm{c}}$, the muon relaxation rates are almost the same under different temperatures and magnetic fields, as expected for muon depolarization by the randomly oriented nuclear dipole moments in the sample. The muon spin relaxation rates are greatly enhanced after entering the superconducting state due to the inhomogenous magnetic field distribution of the flux-line lattice~\cite{Aeppli1987}. The contribution from the flux-line lattice adds in quadrature to the contribution of nuclear dipole moment of the sample:
\begin{eqnarray}
	\label{eq:5}
	\sigma^2=\sigma_{\rm{dip}}^2+\sigma_{\rm{FLL}}^2,
\end{eqnarray}
where $\rm{\sigma_{dip}}$ is the contribution from the randomly oriented nuclear dipolar moments and $\rm{\sigma_{FLL}}$ is the flux-line lattice (FLL) contribution. The temperature dependence of the extracted $\rm{\sigma_{FLL}}$ using Eq.~(\ref{eq:5}) are plotted in Fig.~\ref{fig:Fig3}(b). The behavior that $\sigma$ does not change from the base temperature to $T\sim T_c /3$ suggests the fully gap superconducting gap.

\subsection{\boldmath{C.} Magnetic field and temperature dependence of magnetic penetration depth} 
\label{sec:pd}
 
\begin{figure}[H]
	\begin{center}
		\includegraphics[width=0.45\textwidth]{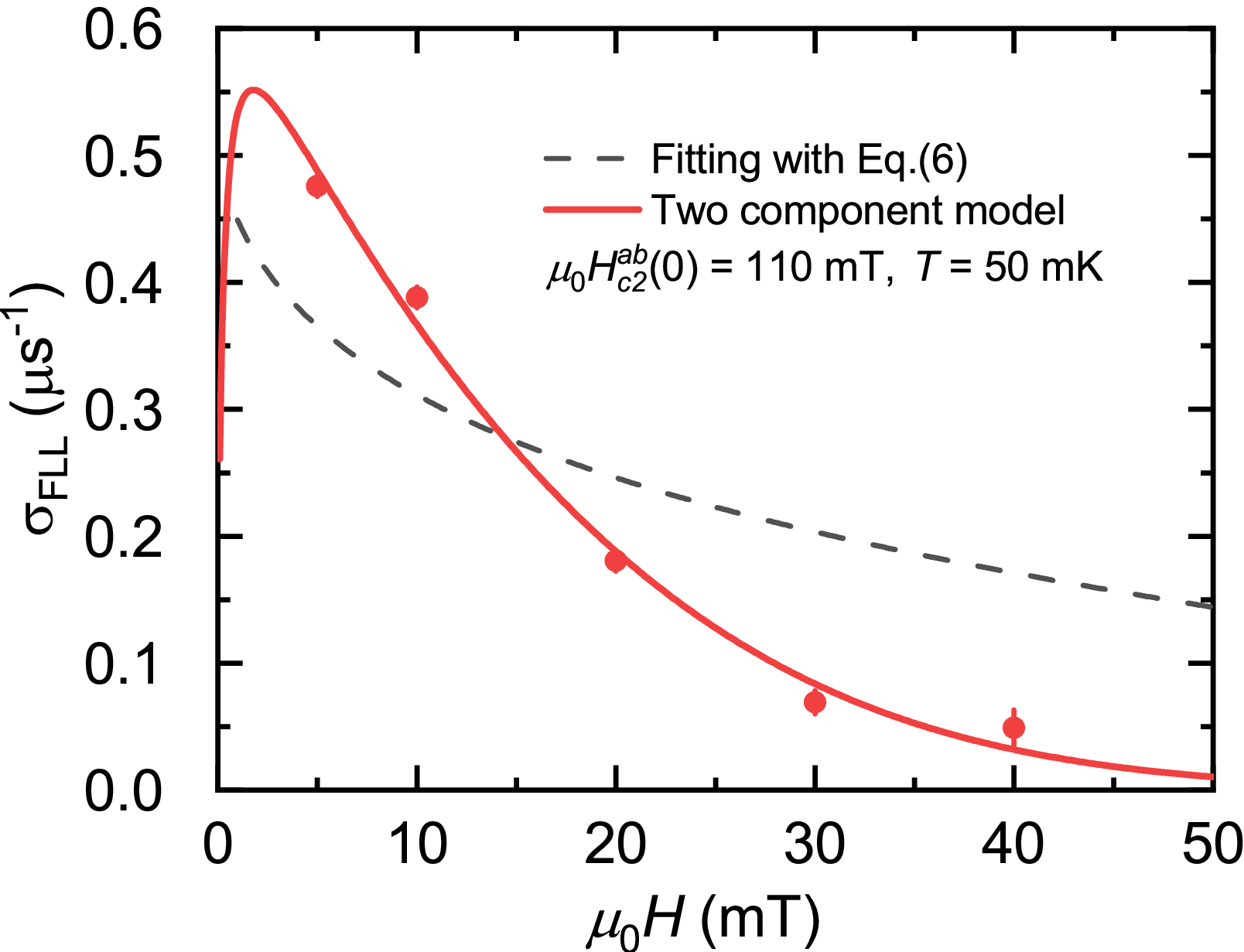}
		\caption{The magnetic-field dependence of muon spin relaxation rate $\sigma_{\rm{FLL}}$. The black dashed line is the theoretically expected $\sigma_{\rm{FLL}}$ with Eq.~(\ref{eq:6}). The red solid line exhibits the fits of data with the two-component Ginzburg-Landau model~\cite{Serventi2004}, using $\xi^{ab}_1=55.1$ nm, $\xi^{ab}_2=71.6$ nm deduced from Ref.~\cite{Ronning2008} and Ref.~\cite{Kurita2009}, resulting $\lambda=0.38(1)$ $\mu$m and $w=0.7(1)$.} 
		\label{fig:Fig4}
	\end{center}
\end{figure}

The internal field distribution of type-$\rm{\rom2}$ superconductors can be described by the penetration depth $\lambda$ and coherence length $\xi$~\cite{Brandt2003}.
For a type-$\rm{\rom2}$ superconductor with $\kappa\geq5$ and a moderate reduced magnetic field $h=H/H_{\rm{c2}}>0.25/\kappa^{1.3}$, $\lambda$ can be calculated with the numerical Ginzburg-Landau model with less than 5\% error~\cite{Brandt2003}:
\begin{multline}
	\label{eq:6}
	\sigma_{\rm{FLL}}[\mu s^{-1}]=4.83\times10^4(1-h)\\\times[1+1.21(1-\sqrt{h})^3]\lambda^{-2}[\rm{nm^{-2}}].
\end{multline}
Noting that the muon spin relaxation rate is nearly temperature independent below $T=0.1$~K, it is reasonable to analyze the field dependence of $\sigma_{\rm{FLL}}$ at base temperature with the zero-temperature upper critical field $H^{ab}_{c2}(0)=0.11$ T taken from Ref~\cite{Kurita2009}. The superscript here indicates that the direction of the external field is in $ab$ plane. For conventional single-band $s$-wave superconductors with medium $\kappa$, the value of $\lambda^{-2}(T)$ and $H^{ab}_{c2}(T)$ can be obtained by fitting the isothermal magnetic field dependence of $\sigma_{\rm{FLL}}$ at different temperatures ~\cite{Barker2015}. The field dependence of $\sigma_{\rm FLL}$ using Eq.~(\ref{eq:6}) with fixed $H^{ab}_{c2}(0)$ at base temperature ($T = 50$ mK) is shown in the black dashed line in Fig.~\ref{fig:Fig4}. It can be seen that the experimentally measured $\sigma_{\rm{FLL}}$ is far from the theoretical calculation, dropping much faster than expected for single-band $s$-wave superconductivity. Noting that the field dependence of $\sigma_{\rm{FLL}}$ described by Eq.~(\ref{eq:6}) is only valid under the assumption of a field-independent $\lambda$~\cite{Brandt2003}, the deviation of $\sigma_{\rm{FLL}}$ from theoretical expectation indicates that $\lambda$ exhibits magnetic field dependence, which is expected for superconductors with nodes or/and multiband superconductors~\cite{Sonier1997,Kadono2004,Weyeneth2010}.
Considering the multiband nature of BaNi$_2$As$_2$~\cite{Subedi2008,Zhou2008}, the field dependence of $\sigma_{\rm{FLL}}$ is further analyzed with the two-component modified London model~\cite{Serventi2004}:
\begin{multline}
	\label{eq:7}
	\overline{\Delta B^2}=\overline{B}^2\sum_{\boldsymbol{q}\neq0}[w\frac{e^{-\boldsymbol{q}^2\xi^2_1/2(1-h_1)}}{1+\boldsymbol{q}^2\lambda^2/(1-h_1)}+\\(1-w)\frac{e^{-\boldsymbol{q}^2\xi^2_2/2(1-h_2)}}{1+\boldsymbol{q}^2\lambda^2/(1-h_2)}],
\end{multline}
where $\boldsymbol{q}=\frac{4\pi}{\sqrt{a}} (\frac{\sqrt3 m}{2}, n+\frac{m}{2})$ are the reciprocal lattice vectors for triangular FLL ($a=1.075\sqrt{\frac{\Phi_0}{B}}$ is the intervortex distance), $\xi_{i}$ is the coherence length for the $i$ th band and $w$ stands for the contribution of the first band to the total superfluid density. The cut off function $e^{-\boldsymbol{q}^2\xi^2/2}$ is a standard approximation for Ginzburg-Landau equations introduced to account for the finite size of vortex cores~\cite{Sonier2007}. Since the four parameters are strongly coupled, it is generally hard to obtain a unique set of solutions. This can be resolved by fixing $\xi_i$ based on the reported properties. The shorter coherence length $\xi^{ab}_1\approx55.1$~nm can be estimated by the upper critical field with $\xi^{ab}_1=(\Phi/2\pi H^{ab}_{c2}(0))^{1/2}$. Here $H^{ab}_{c2}(0)$ is the zero temperature upper critical field for $H \parallel ab$. The ratio of two coherence lengths $\xi^{ab}_1/\xi^{ab}_2=\sqrt{1/0.6}\approx1.3$ can be obtained from the thermal conductivity measurement~\cite{Kurita2009}, which will be discussed later. The fitting result with fixed $\xi^{ab}_{1,2}$ is shown by the solid red line in Fig.~\ref{fig:Fig4},  giving $\lambda=0.38(1)$~$\mu$m and $w=0.7(1)$.

Next, we discuss the temperature dependence of $\lambda^{-2}$. For single-band $s$-wave superconductors with uncertain upper critical fields, $\lambda^{-2}(T)$ and $H^{ab}_{c2}(T)$ can be obtained by fitting the magnetic field dependence of muon relaxation rates under different temperatures~\cite{Barker2015}, while such a method cannot be used when $\lambda$ exhibits magnetic field dependence~\cite{Brandt2003}. We first tried to extract $\lambda(T)$ with Eq.~(\ref{eq:7}). However, the limited data points and the complexity of formula made it hard to get reasonable results in the whole temperature range, especially when the temperatures approached $T_c$. To subtract the $\lambda^{-2}(T)$, Eq.~(\ref{eq:6}) was used under constant magnetic field $H$ with fixed upper critical field $H^{ab}_{c2}(T)$~\cite{Serventi2014}. With the known reduced magnetic field $h$, $\lambda^{-2}(T)$ can be directly calculated from $\sigma_{\rm{FLL}}$.

Considering the two-band feature of \BNA, the temperature dependence of $H^{ab}_{c2}$ was estimated with the empirical two-band model~\cite{Zhu2008,CHANGJAN2011}.
\begin{eqnarray}
	\label{eq:8}
	H^{ab}_{c2}(T)=H^{ab}_{c2}(0)\frac{1-t^{2}}{1+t^{2}},
\end{eqnarray}
where $t=T/T_c$ and $H^{ab}_{c2}(0)=0.11$ T was measured by heat capacity~\cite{Kurita2009}. Since the upper critical fields $H^{ab}_{c2}(T)$ here are estimates, we only calculate $\lambda^{-2}(T)$ at the lowest magnetic field.
 
\subsection{\boldmath{D.} Temperature dependence of superfluid density and gap symmetry} \label{sec:gap-symmetry}
\begin{figure}[H]
	\begin{center}
		\includegraphics[width=0.45\textwidth]{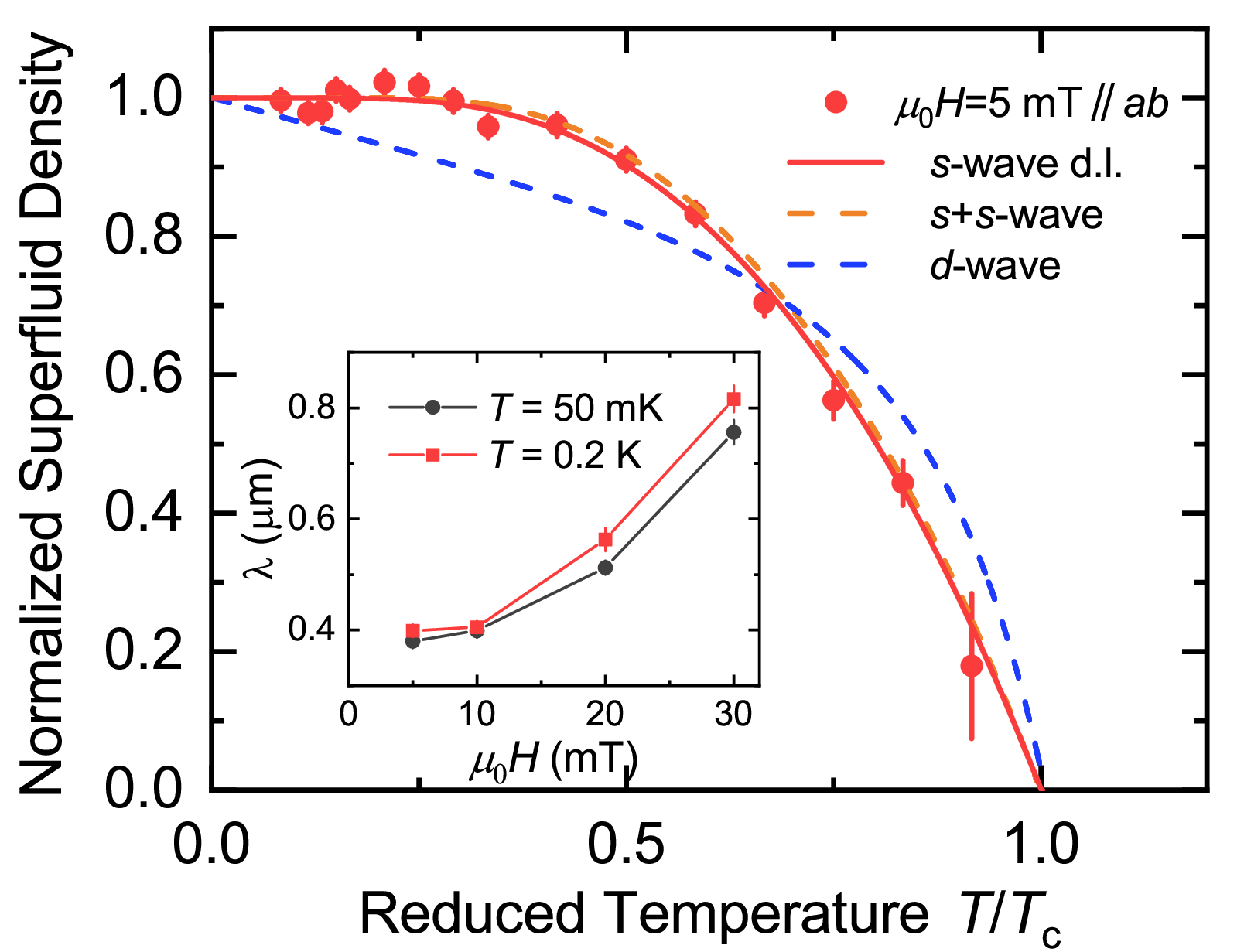}
		\caption{Temperature dependence of the reduced superfluid density versus the reduced temperature $T/T_{\rm{c}}$ measured at $\mu_0H=5$ mT parallel to the $ab$ plane. The colored curves stand for the fitting of Eq.~(\ref{eq:10}) with different gap symmetry functions. The inset shows the field dependence of magnetic penetration depth $\lambda$ at $T=50$ mK and 0.2 K. The connecting lines in the inset are guides for eyes.}
		\label{fig:Fig5}
	\end{center}
\end{figure}

The superfluid density $\rho_{\rm{s}}$ is proportional to $\lambda^{-2}$ based on the London approximation. The resulting normalized superfluid density is plotted in Fig.~\ref{fig:Fig5}. 
To further study the gap symmetry of \BNA, the temperature dependence of $\rho_s$ is fitted with different gap symmetry functions in both dirty limit:
\begin{eqnarray}
	\label{eq:9}
	\frac{\rho_s(T)}{{\rho_s(0)}}=\frac{\Delta(T,\phi)}{\Delta(0,\phi)}{\rm{tanh}}[\frac{\Delta(T,\phi)}{2k_BT}]
\end{eqnarray}
and clean limit,
\small\begin{eqnarray}
	\label{eq:10}
	\frac{\rho_s(T)}{{\rho_s(0)}}=1+\frac{1}{\pi}\int_{0}^{2\pi}d\phi\int_{\Delta(T,\phi)}^{\infty}\frac{\partial f}{\partial E} \frac{E}{\sqrt {E^2-\Delta(T,\phi)^2}}dE,
\end{eqnarray}\normalsize
where $f=[1+\exp(E/k_BT)]^{-1}$ is the Fermi function. $\Delta(T,\phi)=\Delta(\phi)\delta(T/T_c)$ is the gap symmetry function. For $s$-wave model, $\Delta_{\rm{s}}(\phi)=\Delta_0$; for $d$-wave model, $\Delta_{\rm{d}}=\Delta_0\cos(2\phi)$. The temperature dependence $\delta(T/T_c)$ can be approximated as~\cite{Carrington2003}:
\begin{eqnarray}
	\label{eq:11}
	\delta(T/T_c)=\tanh\{1.82[1.018(T_c/T-1)]^{0.51}\},
\end{eqnarray}
where $\rm{\Delta_0}$ is the maximum of zero temperature gap amplitude.  

To examine the possible multi-gap nature inferred from fermi surface structure calculation~\cite{Subedi2008,Zhou2008} and thermal conductivity measurement~\cite{Kurita2009}, we also consider two weakly coupled superconducting bands ($s+s$ or $s+d$), where the linear combination of Eq.(\ref{eq:11}) can be used:
\small\begin{eqnarray}
	\label{eq:12}
	\frac{\lambda^{-2}(T)}{\lambda^{-2}(0)}=w \frac{\lambda^{-2}(T,\Delta_1(T))}{\lambda^{-2}(0,\Delta_1(0))}+(1-w)\frac{\lambda^{-2}(T,\Delta_2(T))}{\lambda^{-2}(0,\Delta_2(0))}.
\end{eqnarray}\normalsize

\begin{table*}[t]
	\centering
		\caption{Fitted parameters to the $\sigma_{\rm{sc}}$ of BaNi$_2$As$_2$ with different models described in the text.}
		\label{table-A}
		\begin{tabular}{>{\centering\arraybackslash} m{0.15\textwidth} >{\centering\arraybackslash} m{0.3\textwidth} >{\centering\arraybackslash} m{0.1\textwidth} >{\centering\arraybackslash} m{0.15\textwidth} >{\centering\arraybackslash} m{0.1\textwidth}}
			\hline
			\hline
			Model & Gap value & $T_{\rm{c}}$ & $2\Delta/k_{\rm{B}}T_{\rm{c}}$ & $\chi^2_{r}$\\
			& (meV) & (K) & \\
			\hline
			$s$ wave c.l. & $\Delta=0.112(3)$ & 0.611(8) & 4.2(1) & 0.98\\
			$s$ wave d.l. & $\Delta=0.101(6)$ & 0.610(8) & 3.8(2) & 1.03\\
			& $\Delta_{\rm1}=0.110(4)$, $\Delta_{\rm2}=0.11(1) $\\
			$s+s$ wave & $w=0.95(51)$ & 0.603(6) & 4.2(4) &1.21  \\
			$d$ wave & $\Delta=0.217(9)$ & 0.574(3) & 8.8(3) & 6.15\\
			\hline
		\end{tabular}
\end{table*}
The fitting results with different models are summarized in Fig.~\ref{fig:Fig5} and Table~\ref{table-A}. The temperature dependence of the normalized superfluid density can be well fitted with both single-gap and multi-gap $s$-wave model. However, the introduction of a second gap does not seem to improve the fitting quality to a great extent, and we can hardly get the second gap value with a reasonable standard error. Therefore, there is no need to introduce more than one gap amplitude to describe $\lambda^{-2}(T)$. The estimated electronic mean free path in the normal state places BaNi$_2$As$_2$ in the dirty limit~\cite{Kurita2009}. The obtained zero temperature superconducting gap  $\Delta_0=0.101(6)$ meV and the critical temperature $T_{\rm{c}}=0.610(8)$ K give $2\Delta/k_{\rm{B}}T_{\rm{c}}=3.8(2)$, which is close to the BCS value of 3.54. The value of superconducting gap is close to  the value of $\Delta_0=0.0946$ meV obtained from heat capacity measurement by Kurita $et$ $al.$~\cite{Kurita2009}.

\subsection{\boldmath{\boldmath{$\rm{\rom4}$}.} DISCUSSION}
\label{sec:discussion}
By analyzing the temperature dependence of penetration depth alone, one may conclude that BaNi$_2$As$_2$ is a common $s$-wave superconductor,  which is consistent with previous heat capacity measurement. However, in single-band $s$-wave superconductors, $\lambda$ should be independent of magnetic field~\cite{Kadono2004,Khasanov2005,Khasanov2008}.  As shown in the inset of Fig.~\ref{fig:Fig5}, $\lambda$ is nearly doubled when the field increased from 5 mT to 40 mT. Such a large change of $\lambda$ cannot be attributed to the error of $H_{c2}(T)$. A field dependence of $\lambda$ is expected for superconductors with nodes or/and multiband superconductors~\cite{Sonier1997,Kadono2004,Weyeneth2010}. The former case can be excluded since the penetration depth below ${T_{c}}/3$ is temperature independent,  which indicates fully-gapped superconductivity. For the latter case, the carriers in one of the bands is suppressed faster than other bands, causing the change of penetration depth. Since $\sigma_{\rm{FLL}}(H)$ can be well described by two-component modified London model, combined with single-gap nature revealed by $\rho_{\rm{s}}(T)$, we suggest that \BNA\ is a multiband single-gap superconductor. The superconducting gap amplitude of each band is nearly the same while different bands exhibit diverse coherence lengths~\cite{Serventi2014}.

A key step in our $\lambda(H)$ analysis is to determine the ratio of different coherence lengths, which is obtained from the thermal conductivity measurement. The shoulder-like anomaly of $\kappa_0(H))/T$ curve was initially attributed to the spread in $H_{\rm{c2}}$~\cite{Kurita2009}, since there was no evidence pointing to a second superconducting band at that time. However, such S-shaped $\kappa_0(H))/T$ curves were later also found in several Ni$_2$X$_2$ based superconductors ~\cite{Kurita2011,Li2014}, which share the same ThCr$_2$Si$_2$ structure and similar electronic structure~\cite{Ronning2009}. Furthermore, the ratio of different coherence lengths in different bands $\xi^{ab}_1/\xi^{ab}_2\approx1.3$ has been confirmed by the calculated band structure~\cite{Shein2009,Subedi2008} and the ARPES experiment~\cite{Zhou2008}. According to the BCS theory, the 
coherence length is proportional to $\left< v_F\right>/\Delta_0$. Here $\left< v_F\right>$ is the averaged value of the Fermi velocity, which can be estimated from the slope of bands crossing the Fermi level.  There are three bands crossing the Fermi level in \BNA. The ratio of the slopes of them is about 1 along $Z$-$A$ and $\sim$ 1-3 along $\Gamma$-$M$~\cite{Shein2009,Subedi2008,Zhou2008}. In consideration of the same $\Delta_0$ of the three bands, $\xi^{ab}_1/\xi^{ab}_2=1.3$ obtained from thermal conductivity data is reasonable.

Considering the uniform gap value in each band revealed by both $\mu$SR and thermodynamic properties~\cite{Kurita2009}, the multiband superconductivity in BaNi$_2$As$_2$ is quite unique compared with other multi-gap superconductors~\cite{Khasanov2009,Khasanov2010,Carrington2003}. It is worth noting that, for a long time, it has been widely believed that the number of excitation gaps determines the number of contributing bands needed in a theoretical model. However, there is no necessary connection between the number of excitation gaps and the number of length scale. Superconductors with multi-bands can be:
i) single-gap and single length scale~\cite{Gupta2020};
ii) multi-gap with single length scale~\cite{Fente2016,Fente2018};
iii) single-gap with multi lengths scales such as SrPt$_3$P~\cite{Khasanov2014} and BaNi$_2$As$_2$ in the present study.
 
In most cases, strong electronic scattering in conventional dirty superconductors weakens the difference between bands. As a result, each band tends to exhibit identical superconducting behavior. A natural question here is what causes the unconventional multiband behavior in BaNi$_2$As$_2$.  Recent density functional theory (DFT) results showed that the rebonding of As anions can induce complex structural instabilities. These low-temperature structures are similar in energy but are very different in Fermiology~\cite{Lei2022}. The competition between these bonding arrangements can possibly explain this unconventional superconducting behavior. However, such scenario is hard to reconcile with the single gap amplitude revealed by heat capacity~\cite{Kurita2009} and $\mu$SR experiments. On the other hand, recent Raman scattering measurements revealed a strong dynamic nematic fluctuations in BaNi$_2$As$_2$, which strongly supported the nematic enhanced superconductivity upon doping in this material~\cite{Yao2022}. Whether the unconventional single-gap multiband superconducting in BaNi$_2$As$_2$ is a manifestation of the interaction between superconducting and nematicity demands further confirmation. Subsequent researches on superconducting band properties in doped systems will be very helpful to elucidate how the CDW and charge induced nematicity interact with superconductivity.

\section{\boldmath{$\rm{\rom5}$}. CONCLUSIONS} \label{sec:conc}
To summarize, systematic $\mu$SR experiments have been performed to study the temperature and magnetic field dependence of penetration depth $\lambda$ in BaNi$_2$As$_2$. The temperature independent behavior of superfluid density $\rho_{\rm{s}}$ at low temperature suggests its fully-gapped nature. The fitting with single-gap $s$-wave model in dirty limit gives the zero temperature gap amplitude $\Delta_0=0.101(6)$ meV.  The deviation of $\sigma_{\rm{FLL}}(H)$ from single-band scenario can be understood with a two-component model with different coherence lengths in each band. Our TF-$\mu$SR results reveal that BaNi$_2$As$_2$ is a rare single-gap multiband superconductor. The ZF-$\mu$SR experiments confirmed the preservation of time-reversal symmetry in \BNA. This work provides new experimental evidence to help understand the interplay between CDW, nematicity and superconductivity.

\section{ACKNOWLEDGMENTS} \label{sec:ack}
We are grateful to the ISIS cryogenics Group for their valuable help during the $\mu$SR experiments (ISIS.RIB 2010283). This research was funded by the National Natural Science Foundations of China, No.~12174065, the National Key Research and Development Program of China, No.~2022YFA1402203, and the Shanghai Municipal Science and Technology Major Project Grant, No.~2019SHZDZX01. The materials synthesis efforts at Rice are supported by the U.S. Department of Energy (DOE), Basic Energy Sciences (BES), under Contract No. DE-SC0012311, and the Robert A. Welch Foundation, Grant No. C-1839 (P.D.).

\appendix
\renewcommand\thefigure{\Alph{section}\arabic{figure}}  
\section{APPENDIX: CHARACTERIZATION MEASUREMENTS ON SAMPLES POSSIBLY EXPOSED TO THE AIR}
\newcounter{Afigure}
\setcounter{Afigure}{1}
\renewcommand{\thefigure}{A\arabic{Afigure}}   
\BNA\ samples are air sensitive~\cite{Ronning2008}. After the $\mu$SR experiments, single crystals of \BNA\ were stored in a glass bottle filling with nitrogen, and were shipped from the $\mu$SR facility to our university. Due to the epidemic prevention policy, our sample got stuck in customs for several months. The single crystal XRD and electrical resistance experiments were performed on samples which were possibly exposed to the air for some time.

\begin{figure}[H]
	\centering
	\includegraphics[width=0.48\textwidth]{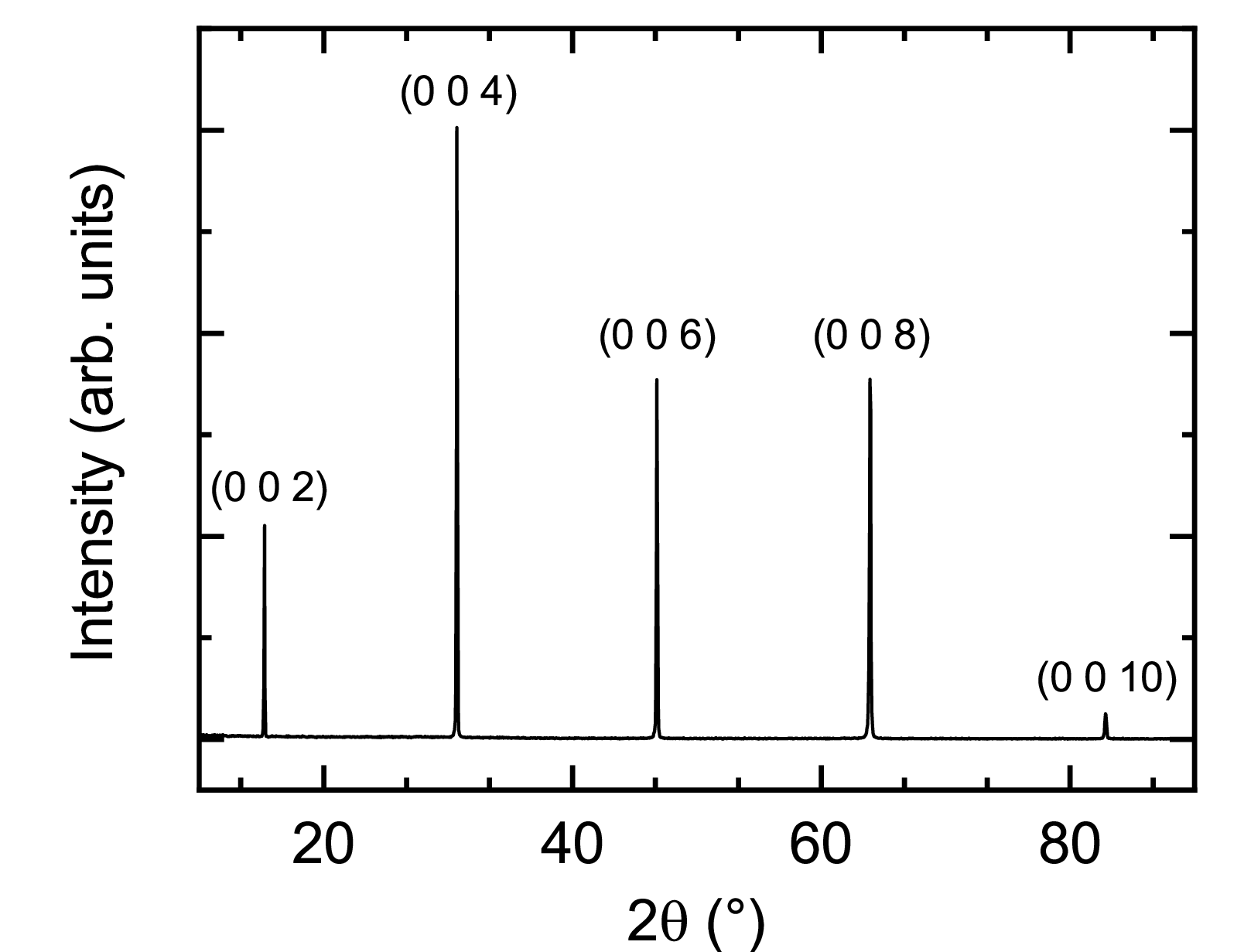}
	\caption{Single crystal X-ray diffraction patterns of BaNi$_2$As$_2$.}
	\label{fig:FigA1}
	\addtocounter{Afigure}{1}
\end{figure}

The single crystal XRD diffraction patterns were obtained by using a Bruker D8 advanced X-ray diffraction spectrometer ($\lambda=1.5418\ \rm{\r{A}}$) at room temperature. Single crystal \BNA\ are naturally  oriented with the $c$-axis normal to the plate, and the XRD patterns are shown in Fig.~\ref{fig:FigA1}. The diffraction peaks of specific lattice planes are obvious. The lattice parameter $c$ is determined via the liner fitting between the diffraction pattern index $n$ and $\sin(\theta)$, which gives
$c$ = 11.66(4) $\rm{\r{A}}$. For comparison, $c$ = 11.54(2) $\rm{\r{A}}$ in Ref~\cite{Ronning2008}.

\begin{figure}[H]
	\centering
	\includegraphics[width=0.48\textwidth]{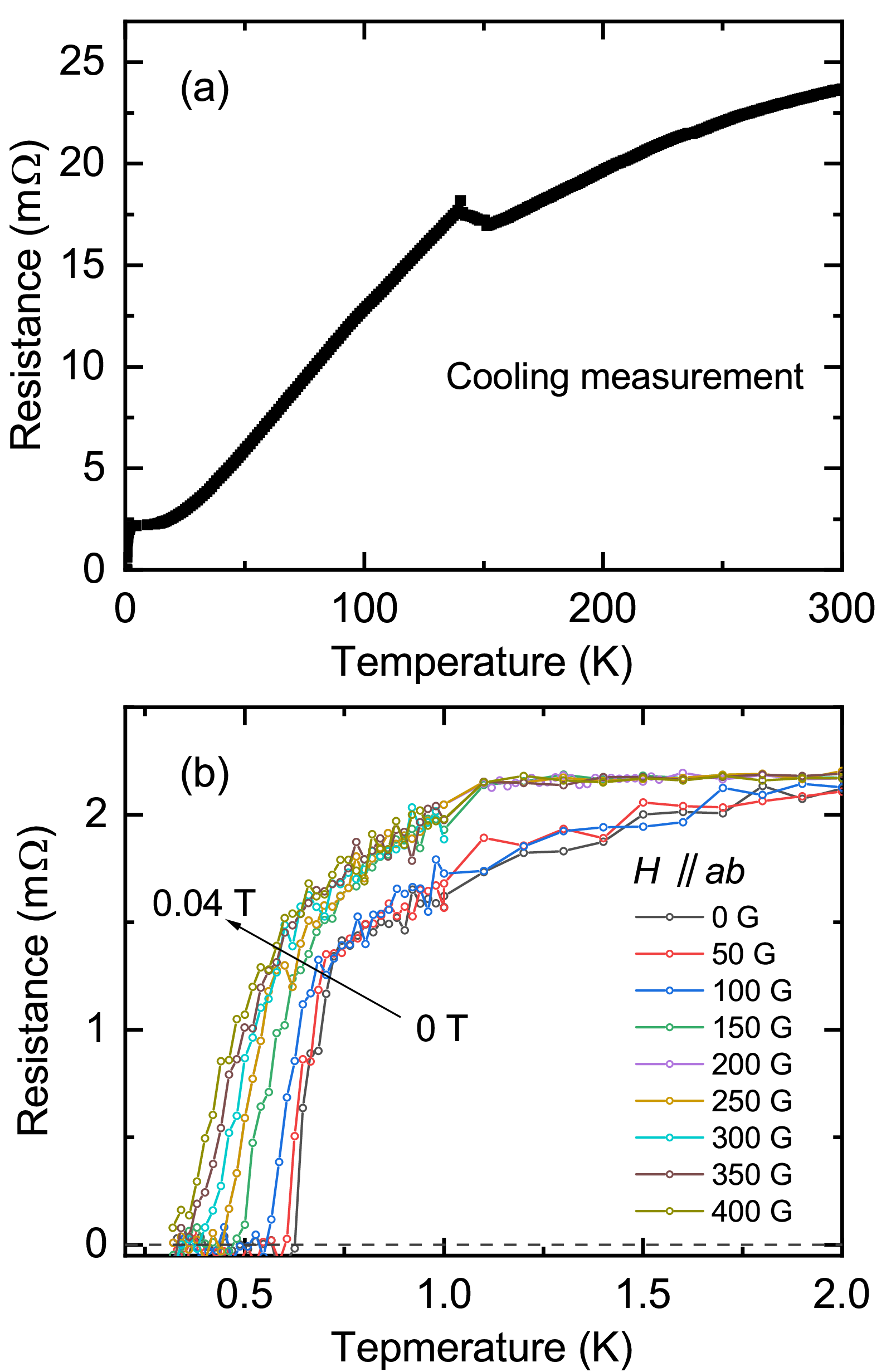}
	\caption{Temperature dependence of resistance from 300 K to 2 K(a) , and from 2.0 K to 0.3 K (b) with external magnetic field parallel to $ab$ plane. There is an additional 10 minutes' exposure to air between the 100 G and 150 G measurements in (b).}
	\label{fig:FigA2}
	\addtocounter{Afigure}{1}
\end{figure}

Temperature dependence of resistance $R(T)$ from 300 K to 0.3 K was measured using standard four-probe method on a physical property measurement system (PPMS) equipped with Helium3 refrigerator. The resistance was measured with cooling. Fig.~\ref{fig:FigA2}(a) shows the temperature dependence of resistance $R(T)$ from 300 K to 2 K. A broad hump is observed near 150 K, which is identified as the structural phase transition~\cite{Ronning2008}. The width of transition is much broader compared to Ref~\cite{Ronning2008}, which may due to the damage of our sample after its exposure to the air. The RRR (=$\rho$(300 K)/$\rho$(4 K)) is 10. For comparison, RRR = 5 in Ref~\cite{Ronning2008}. The resistance below 2 K is shown in Fig.~\ref{fig:FigA2}(b). The direction of the external magnetic field is parallel to $ab$-plane. The resistance of the sample reaches zero at 0.63 K under zero field and the sample shows superconductivity under the external field of 400 G. It should be noted that, the measurements with field higher than 100 G was conducted after the sample being exposed to the air for another 10 minutes. There are obvious differences between $R-T$ curves over and below 100 G, which further illustrate that \BNA\ is very air-sensitive.

\end{document}